\documentclass[aps,prc,reprint,superscriptaddress,floatfix,amsmath]{revtex4-2}
\usepackage{graphicx,epstopdf}
\usepackage{booktabs}
\usepackage{multirow}

\begin{document}

\graphicspath{{./}{figs}}

\title{Exploration for Astromers near $^{132}$Sn with the Canadian Penning Trap}

\author{A.A. Valverde}
\email[Corresponding author: ]{avalverde@anl.gov}
\affiliation{Physics Division, Argonne National Laboratory, Lemont, IL 60439, USA }

\author{S. Cupp}
\affiliation{Theoretical Division, Los Alamos National Laboratory, Los Alamos, NM 87545, USA}
\affiliation{Center for Theoretical Astrophysics, Los Alamos National Laboratory, Los Alamos, NM 87545, USA}

\author{A. Gross}
\affiliation{Theoretical Division, Los Alamos National Laboratory, Los Alamos, NM 87545, USA}
\affiliation{Center for Theoretical Astrophysics, Los Alamos National Laboratory, Los Alamos, NM 87545, USA}

\author{B. Liu}
\affiliation{Physics Division, Argonne National Laboratory, Lemont, IL 60439, USA }
\affiliation{Department of Physics and Astronomy, University of Notre Dame, Notre Dame, IN 46556, USA}

\author{M.R. Mumpower}
\affiliation{Obsidian Research, Fort Wayne, IN 46835, USA}
\affiliation{Department of Physics and Astronomy, University of Notre Dame, Notre Dame, IN 46556, USA}
\affiliation{Computational Division, Los Alamos National Laboratory, Los Alamos, NM 87545, USA}
\affiliation{Center for Theoretical Astrophysics, Los Alamos National Laboratory, Los Alamos, NM, 87545, USA}

\author{G.W. Misch}
\affiliation{X Theoretical Design, Los Alamos National Laboratory, Los Alamos, NM 87545, USA}

\author{W.S. Porter}
\affiliation{Department of Physics and Astronomy, University of Notre Dame, Notre Dame, IN 46556, USA}

\author{D. Ray}
\altaffiliation[Present address: ]{TRIUMF, Vancouver, BC V6T 2A3, Canada}
\affiliation{Physics Division, Argonne National Laboratory, Lemont, IL 60439, USA }
\affiliation{Department of Physics and Astronomy, University of Manitoba, Winnipeg, MB R3T 2N2, Canada}

\author{M. Brodeur}
\affiliation{Department of Physics and Astronomy, University of Notre Dame, Notre Dame, IN 46556, USA}

\author{D.P. Burdette}
\affiliation{Physics Division, Argonne National Laboratory, Lemont, IL 60439, USA }

\author{N. Callahan}
\affiliation{Physics Division, Argonne National Laboratory, Lemont, IL 60439, USA }

\author{A. Cannon}
\altaffiliation[Present address: ]{TRIUMF, Vancouver, BC V6T 2A3, Canada; Department of Physics and Astronomy, University of Victoria, Victoria, BC V8W 3P6, Canada}
\affiliation{Department of Physics and Astronomy, University of Notre Dame, Notre Dame, IN 46556, USA}

\author{J.A. Clark}
\affiliation{Physics Division, Argonne National Laboratory, Lemont, IL 60439, USA }

\author{A.T. Gallant}
\affiliation{Nuclear and Chemical Sciences Division, Lawrence Livermore National Laboratory, Livermore, CA 94550, USA}

\author{D.E.M. Hoff}
\affiliation{Nuclear and Chemical Sciences Division, Lawrence Livermore National Laboratory, Livermore, CA 94550, USA}

\author{A.M. Houff}
\affiliation{Physics Division, Argonne National Laboratory, Lemont, IL 60439, USA }
\affiliation{Department of Physics and Astronomy, University of Notre Dame, Notre Dame, IN 46556, USA}

\author{K. Kolos}
\affiliation{Nuclear and Chemical Sciences Division, Lawrence Livermore National Laboratory, Livermore, CA 94550, USA}

\author{F.G. Kondev}
\affiliation{Physics Division, Argonne National Laboratory, Lemont, IL 60439, USA }

\author{O.S. Kubiniec}
\affiliation{Physics Division, Argonne National Laboratory, Lemont, IL 60439, USA }

\author{A. LaLiberte}
\affiliation{Physics Division, Argonne National Laboratory, Lemont, IL 60439, USA }
\affiliation{Department of Physics, University of Chicago, Chicago, IL 60637, USA }

\author{G.E. Morgan}
\altaffiliation[Present address: ]{Cyclotron Institute, Texas A\&M University, College Station, TX 77843, USA}
\affiliation{Physics Division, Argonne National Laboratory, Lemont, IL 60439, USA }
\affiliation{Department of Physics and Astronomy, Louisiana State University, Baton Rouge, LA
70803, USA}

\author{R. Orford}
\affiliation{Nuclear Science Division, Lawrence Berkeley National Laboratory, Berkeley, CA 94720, USA}

\author{C. Quick}
\altaffiliation[Present address: ]{Department  of Physics and Astronomy, University of Tennessee, Knoxville, TN 37996, USA}
\affiliation{Department of Physics and Astronomy, University of Notre Dame, Notre Dame, IN 46556, USA}

\author{F. Rivero}
\affiliation{Department of Physics and Astronomy, University of Notre Dame, Notre Dame, IN 46556, USA}

\author{D. Santiago-Gonzalez}
\affiliation{Physics Division, Argonne National Laboratory, Lemont, IL 60439, USA }

\author{G. Savard}
\affiliation{Physics Division, Argonne National Laboratory, Lemont, IL 60439, USA }
\affiliation{Department of Physics, University of Chicago, Chicago, IL 60637, USA }

\author{N.D. Scielzo}
\affiliation{Nuclear and Chemical Sciences Division, Lawrence Livermore National Laboratory, Livermore, CA 94550, USA}

\author{K.S. Sharma}
\affiliation{Department of Physics and Astronomy, University of Manitoba, Winnipeg, MB R3T 2N2, Canada}

\author{L. Varriano}
\altaffiliation[Present address: ]{Center for Experimental Nuclear Physics and Astrophysics, University of Washington, Seattle, WA 98195, USA}
\affiliation{Physics Division, Argonne National Laboratory, Lemont, IL 60439, USA }
\affiliation{Department of Physics, University of Chicago, Chicago, IL 60637, USA }

\date{\today}

\begin{abstract}
Nuclear isomers can have significant impacts on astrophysical nucleosynthesis processes, with recent efforts demonstrating that the population of isomeric states with different half-lives may require separate treatment in reaction networks to accurately capture the differences in heating or in identifiable electromagnetic signals. Several potential so-called ``astromers'' in tin and antimony isotopes near doubly-magic $^{132}$Sn were identified and direct mass measurements of their ground and isomeric states were performed with the Canadian Penning Trap at Argonne National Laboratory's CARIBU facility, and their impact on astrophysical reaction rates and in reaction networks calculated. It was found that $^{129g,m}$Sn, with measured mass excesses of $-80 593.2(25)$ keV and $-80 557.4(25)$ keV, respectively, and an excitation energy of $35.8(35)$ keV, behaves as an astromer during neutron capture in the $i$-process and in the $r$-process. 
\end{abstract}

\pacs{}
\maketitle

\section{Introduction}
The synthesis of the vast majority of isotopes heavier than iron occurs in the hot, neutron-rich environments where the rapid ($r$), intermediate ($i$), and slow ($s$) neutron capture processes occur. Our understanding of these astrophysical processes relies in part on models employing nuclear reaction networks. These reaction networks incorporate various pieces of nuclear data, including masses, half-lives, and capture rates. Excited nuclear states are traditionally treated either by only considering the ground state or by considering a Boltzmann distribution of all excited nuclear states. It has long been recognized that a species with an isomeric state can fail to reach thermal equilibrium, and so both ground and isomeric states must be treated as separate nuclear species in a network, such as in the notable case of $^{26}$Al \cite{Ward80,Prantzos96,Runkle01,Iliadis11}.

Recently, there has been renewed interest in the impacts of isomeric states on nucleosynthesis in astrophysical environments. Isomeric states can have significantly different decay lifetimes, among other properties, from their ground state counterparts, and can affect nucleosynthesis processes through deferring or accelerating heating by impacting $\beta$-decay lifetimes or produce identifiable electromagnetic signals \cite{Misch21}. Recent studies have explored the impact of isomers on $r$-process simulations that are populated either directly in decays or thermally in the astrophysical environment \cite{Fujimoto20,Misch21,Misch24}. In the $s$ process, studies have also considered isomers populated during neutron capture \cite{Tannous25}.

Such astrophysically relevant isomers are called ``Astromers'' \cite{Misch21}. Recent experimental efforts have shown the impact that new, high precision mass measurements can have on our understanding of these states \cite{Hoff23,Rivero25}, and studies of the impact of potential astromers has shown the presence of several candidates in the vicinity of doubly-magic $^{132}$Sn \cite{Misch21}. We thus embarked on a campaign of Penning trap mass measurements of potential astromers in antimony and tin isotopes in this vicinity to determine isomeric excitation energies and mass excess of these nuclei.

\section{Experimental Setup and Analysis}
Here we report new mass values for neutron-rich tin and antimony isotopes near $^{132}$Sn produced at Argonne National Laboratory's CAlifornium Rare Isotope Breeder Upgrade (CARIBU) facility~\cite{Savard08,Savard11} and measured with the Canadian Penning Trap mass spectrometer (CPT)~\cite{Savard01}. 
CARIBU beam production begins by slowing fission fragments from a $\sim0.5$ Ci $^{252}\text{Cf}$ source with a thin gold foil and then stopping them in a large-volume, helium-filled gas catcher. 
After the beam is extracted through a radiofrequency quadrupole (RFQ), it passes through a high-resolution magnetic mass separator \cite{Davids08} which provides preliminary mass separation based on mass number over charge ($A/q$).
This continuous beam is then injected into an RFQ cooler-buncher, reducing the emittance of the beam via collisions with moderate-pressure helium gas, and accumulating and releasing the beam in bunches every 50 ms.
The bunched ions then enter the multi-reflection time-of-flight mass separator (MR-TOF) \cite {HIRSH16} where they remain for approximately 15 ms, allowing for the separation of the ions of interest from isobaric contaminants with a resolving power of $\frac{m}{\Delta m} > 10^5$. The ions of interest are then selected with a Bradbury-Nielsen gate \cite{Bradbury36} at the exit of the MR-TOF before delivery to the CPT experimental setup. On entering the CPT experimental setup, ions are captured and further cooled in a linear RFQ trap before injection into the CPT.
The calibrant species used for these measurements, $^{133}$Cs$^{+}$, is instead produced using the CPT's Stable Ion Source, a thermal alkali source located just upstream of the linear RFQ trap.

\begin{figure}[t!]
\centering
 \includegraphics[width=\columnwidth]{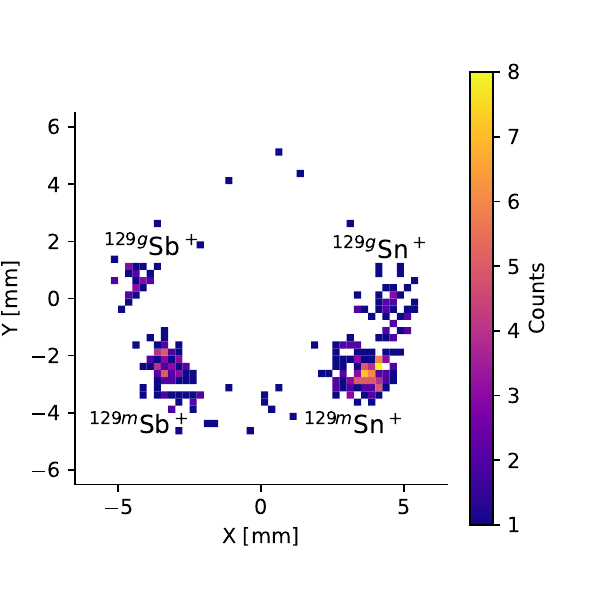}
 \caption{A histogram of detected ion locations for a sample $^{129g}\text{Sn}^+$ final phase measurement with phase accumulation time $t_{\text{acc}}=450.894$ ms. The locations of the four species captured in the trap, which separate due to differing mass-based phase accumulations, are labeled. \label{fig:1Spots}}
\end{figure}

Once ions are captured in the Penning trap, the Phase-Imaging Ion-Cyclotron-Resonance (PI-ICR) technique \cite{Eliseev13} is used to measure the cyclotron frequency. As detailed in Ref.~\cite{Orford20}, in the implementation used at the CPT, the cyclotron frequency ($\nu_c$) of an ion is determined through the simultaneous measurement of the reduced cyclotron frequency and the magnetron frequency. 
For measurements, ions are kept in the trap for a fixed amount of time $T$. They are first placed into their reduced cyclotron eigenmotion, where they are allowed to accumulate phase for some time $t_{\text{acc}}$, and then a quadrupole RF pulse at approximately $\nu_c$ is applied on the Penning trap's segmented ring electrode to convert this to the magnetron eigenmotion, where they accumulate phase for time $T-t_{\text{acc}}$. Two classes of measurements are taken, reference phase measurements where $t_{\text{acc}}=0$ and only magnetron phase is accumulated, and final phase measurements where $t_{\text{acc}}>0$ and both magnetron and reduced cyclotron phase are accumulated. The cyclotron frequency can then be determined from the difference between these two phases $\phi_c$ by:
\begin{equation}
\nu_c=\frac{\phi_{\text{tot}}}{2\pi t_{\text{acc}}}=\frac{\phi_c+2\pi N}{2\pi t_{\text{acc}}},
\end{equation}
where $\phi_{\text{tot}}$ is the total phase accumulated and $N$ is the number of complete revolutions the ion undergoes in $t_{\text{acc}}$. 
Several shorter $t_{\text{acc}}$ measurements, from single to hundreds of ms, are used to identify all of the beam components, including isomeric states. A series of longer $t_{\text{acc}}$ measurements is then done to produce a final measurement, following the techniques described by Ref. ~\cite{Orford20} to minimize potential systematic uncertainties. 
A sample final phase measurement with $t_{\text{acc}}=450.894$ ms of $^{129g}\text{Sn}^+$ with all the visible beam components identified can be seen in Fig.~\ref{fig:1Spots}; spots were then clustered and analyzed using a Gaussian mixture model code~\cite{Weber22}.

\begin{figure}[t!]
\centering
 \includegraphics[width=\columnwidth]{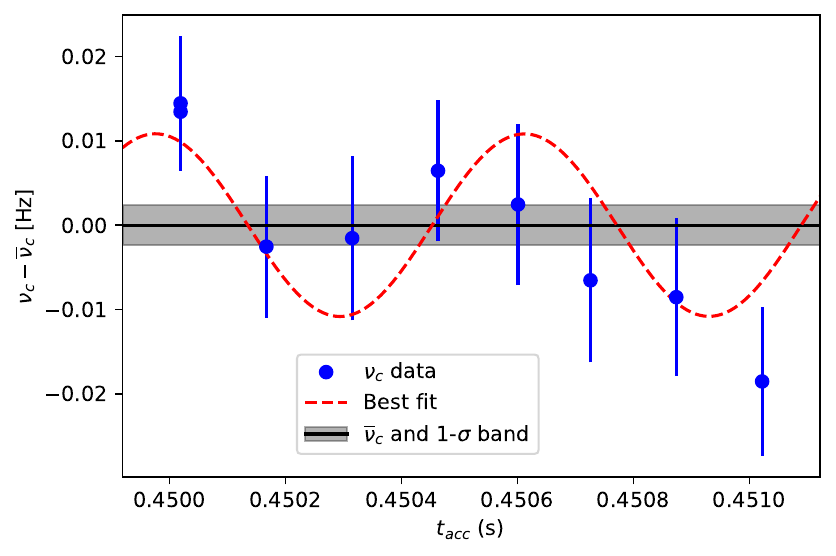}
 \caption{Measured $\nu_c$ values for $^{129m}\text{Sn}^+$ at eight distinct $t_{\text{acc}}$ values between 450.019 ms and 451.022 ms. The dashed line represents a fit of the model described in Ref. \cite{Orford20} to the data, and the solid line and bar the true $\bar{\nu}_c$.  \label{fig:2Sine}}
\end{figure}

As described in Ref. \cite{Orford20}, there are two significant sources of potential systematic uncertainty to be corrected for in these PI-ICR measurements. The first can be seen as a clear $t_{\text{acc}}$-dependent oscillation in measured $\nu_c$, which is well-described as a sinusoidal oscillation at the magnetron frequency around the true cyclotron frequency $\bar{\nu}_c$. Thus, a series of $\nu_c$ measurements are taken at various $t_{\text{acc}}$ across roughly one and a half magnetron periods and then fit using the model described in Ref. \cite{Orford20} to find $\bar{\nu}_c$. A sample plot of eight final phase measurements taken for $t_{\text{acc}}$ values between 450.019 and 451.022 ms and the fit to determine $\bar{\nu}_c$ for $^{129m}$Sn can be seen in Fig. \ref{fig:2Sine}. 

The other significant correction occurs in the reference phase measurement due to the small but significant accumulation of mass-dependent phase during the excitation periods. Following the procedure in Ref. \cite{Orford20}, this is corrected using an iterative process where the reference phase correction is recalculated using the newly-determined cyclotron frequency until the change in the correction is more than an order of magnitude smaller than the statistical uncertainty. For these data, this correction was smaller than a part in $10^9$.

The experimental result in Penning trap mass spectrometry is the frequency ratio between the $\nu_c$ of the isotope of interest and that of a well-known calibrant mass used to determine the magnetic field strength. For all these measurements, $^{133}$Cs was used as the calibrant mass, and so for a given isotope $^{A}$X, 
\begin{equation}
R=\bar{\nu}_c(^{133}\text{Cs}^+)/\bar{\nu}_c(^{A}\text{X}^+).
\end{equation}
The effect of the non-circular projection from the trap to the position sensitive MCP has been corrected using the method described in \cite{Liu25}, and the effect of ion-ion interactions has been reduced by limiting analysis to events with 5 or fewer ions detected. Most remaining systematic uncertainties, arising from sources like magnetic field inhomogeneities, misalignment of the trap electrodes, and non-harmonic terms in the trap potential should scale linearly with the difference in mass over charge between the ion of interest and the calibrant. These have been studied using well-known stable isotopes and found in this mass region to have an effect of 4.1 $\times10^{-10}$/$\Delta(A/q)$ \cite{Liu25,Ray24}, which was applied and added in quadrature to the statistical uncertainty on $R$. Remaining systematic uncertainties associated with temporal instabilities in the magnetic field, the electric field in the Penning trap, ion-ion interactions, and the non-circular final projection of the phase have been studied~\cite{Ray25} and determined to have a cumulative effect smaller than 4 parts in $10^9$; this was added in quadrature. 
The mass can then be determined via,
\begin{equation}
    M(^{A}\text{X})=[M(^{133}\text{Cs})-m_e]R+m_e,
\end{equation}
where $m_e$ is the electron mass.
A summary of masses measured in this work and a comparison with previously measured values is presented in Table~\ref{tab:mass}.

\begin{table*}[t!]
 \caption{\label{tab:mass} Spins and parities (I$^{\pi}$), half-lives (T$_{1/2}$), cyclotron frequency ratios ($R=\bar{\nu}_c(^{133}\text{Cs}^+)/\bar{\nu}_c(^A\text{X}^+)$), mass excesses (\emph{ME}), and excitation energies $E_x$ for the ground and isomeric states of nuclides ($X$) measured in the present work, compared to literature values, with \# indicating an estimated value from the AME or NUBASE based on trends in the mass surface.}
\begin{ruledtabular}
\begin{tabular}{c c c c c c c c c c}
       & & & \multicolumn{3}{c}{Present work}  & \multicolumn{4}{c}{Literature values [keV]} \\
\cmidrule(lr){4-6} \cmidrule(lr){7-10}
Nuclide ($X$) & $I^\pi$ &$T_{1/2}$ [m] &$R$ & $ME$ [keV] & $E_x$ [keV] & $ME$$^{a)}$ & $E_x$$^{a)}$  & $ME$ & $E_x$ \\ 
\hline
$^{129g}$Sn & $3/2^+$ & 2.23(4)& 0.969 963 694(20) & -80 593.2 (25) &        & -80 591(17)& & & \\
$^{129m}$Sn & $11/2^-$ & 6.9(1)& 0.969 963 983(20) & -80 557.4 (25) & 35.8(35) & -80 556(17)& 35.15(5) & &\\
$^{131g}$Sn & $3/2^+$ & 0.933(8)& 0.985 038 825(16) & -77 278.2 (19) &        & -77 265(4) & & &\\
$^{131m}$Sn & $11/2^-$ & 0.973(8)& 0.985 039 356(15) & -77 212.4 (18) & 65.8(26)& -77 200(4) & 65.1(3) & & 64(2)$^{b)}$\\
$^{132g}$Sb & $(4)^+$ & 2.79(7)& 0.992 543 964 0(82) & -79 635.1 (10) &       & -79 635.3(25) & & -79 629.6(27)$^{c)}$& \\
$^{132m}$Sb & $(8^-)$ & 4.10(5)& 0.992 545 137 8(88) & -79 489.8 (11) & 145.3(15)& -79 490(50)$\#$& 150(50)$\#$ & -79 490.3(33)$^{c)}$& 139.3(20)$^{c)}$\\
\end{tabular}
\end{ruledtabular}
$^{a)}$AME2020 and NUBASE2020~\cite{AME20,NUBASE20}.;
$^{b)}$PI-ICR in Benito \emph{et al}.~\cite{Benito24}.
$^{c)}$Jaries \emph{et al}.~\cite{Jaries24}.
\end{table*}

\section{Results}
\subsection{$^{129g,m}$Sn}
Our new results for $^{129g,m}$Sn agree with the values from AME2020 \cite{AME20}, but are over 6 times more precise. The previous value comes from a Time-of-Flight Ion-Cyclotron-Resonance (TOF-ICR) Penning trap measurement that could not resolve the isomeric and ground states by Sikler \emph{et al.} \cite{Sikler05} (36\% contribution to the AME Value), and from $\beta$-endpoint measurements of $^{129}\text{In}(\beta^-)^{129}\text{Sn}$ (44\%) and $^{129m}\text{In}(\beta^-)^{129}\text{Sn}$ (20\%) \cite{Gausemel04}. Our direct measurement of the isomeric state gives an excitation energy that is in good agreement with the value quoted in NUBASE2020 \cite{NUBASE20}.

\subsection{$^{131g,m}$Sn}   
Our new results for $^{131g,m}$Sn are a factor of 2 more precise than the AME2020 values \cite{AME20}. Our values are approximately $3\sigma$ lighter than the AME value, which comes primarily from two previous TOF-ICR Penning trap measurements. Dworschak \emph{et al.} \cite{Dworschak08} could not resolve the ground and isomeric states, while Van Schelt \emph{et al.} \cite{VanSchelt13,VanSchelt12} did. Our results agree with Van Schelt \emph{et al.} for the isomeric state very well, and our measured excitation energy is in good agreement with the value from NUBASE2020 \cite{NUBASE20} as well as the excitation energy directly measured using the PI-ICR technique at JYFLTRAP \cite{Benito24}. We note that in production from spontaneously-fissioning $^{252}$Cf, the isomeric state is more populated than the ground state. 

\subsection{$^{132g,m}$Sb}
Our new result for $^{132g}$Sb is better than a factor of two more precise than the AME2020 value \cite{AME20}, and our value for $^{132m}$Sb is in good agreement with the estimate from NUBASE2020 \cite{NUBASE20}. The previous ground state mass excess for this nucleus comes from other Penning trap mass measurements, with 83\% coming from a TOF-ICR measurement by Hakala \emph{et al.,} \cite{Hakala12} and 17\% from a TOF-ICR measurement by Van Schelt \emph{et al.} \cite{VanSchelt13}, both of which our measurement is in good agreement with. Our isomeric state is in good agreement with the value from Van Schelt \emph{et al.} as well \cite{VanSchelt12}, as is our measured excitation energy, but both are a factor of 15 more precise. While Kankainen \emph{et al.}\cite{Kankainen13} did not produce a definite value for this excitation energy, we are in agreement with their preliminary value of 153(14) keV. Jaries \emph{et al.} have also recently measured this isotope \cite{Jaries24}; while our measured isomeric states are in good agreement, their ground state mass excess disagrees with previous work by a little over 2$\sigma$, and thus their excitation energy is also in disagreement with ours.

\section{Discussion}

We now explore the impact of the new experimental measurements in the context of the astrophysical nucleosynthesis processes.  
The new mass measurements enter into the calculation of astrophysical reaction rates, via the thermal population of the excited states, as well as in their respective decay \cite{Clark23}. 
These nuclei may be produced in both the $i$-process and the $r$-process. 
Since the astrophysical conditions associated with the $i$ process and $r$ process differ, they offer complementary avenues for exploring the astromeric character of these nuclei \cite{Misch24}. 

An isomer is classified as an astromer when the timescale of internal transitions is greater than the timescales of its nuclear destruction channels. 
Quantitatively, this results in the following inequalities for the rates: 

\begin{align}
\lambda_{gs \to m} &< \sum \lambda_{\text{destruction}, gs} \label{eqn:iso_dest_gs}\\
\lambda_{m \to gs} &< \sum \lambda_{\text{destruction}, m}
\label{eqn:iso_dest_m}
\end{align}

If either inequality is satisfied, the isomer cannot maintain thermal distribution and will behave as a metastable species in the astrophysical environment. 

An astromer is influential if it impacts the nucleosynthesis or an observable quantity. Quantitatively, one measure of this influence is the Astromer Importance Rating (AIR) metric developed by Misch and colleagues in Ref. \cite{Misch21}.
When this number is relatively large the isomer may be influential and should be examined. 
When the value is small the isomer has little to no influence. 

The transition between an isomer behaving as a distinct species and being in thermal equilibrium with the ground state is characterized by the thermalization temperature. 
We define the temperature at which the internal equilibration rate between the ground state and the isomer ($\lambda_{gs \leftrightarrow m}$) first exceeds the total (summed) nuclear destruction rate (predominantly $\beta$-decay in the environments considered here); recall the above equations. 

At temperatures below the thermalization temperature, the barriers to internal transitions are too high for the thermal photon bath to overcome on the timescale of the nuclear transmutations. 
In this regime, the population of the isomer and ground state must be treated as separate evolutionary species. 
Above the thermalization temperature, the internal transitions are sufficiently rapid to maintain a Boltzmann distribution between the levels. 
It is important to note that the thermalization is not simply a function of the isomer’s excitation energy. 
Rather, it depends on the existence of intermediate states that connect the ground state and isomer. 
If low-lying intermediate states with strong transition strengths exist, thermalization can occur at temperatures significantly lower than the isomer's excitation energy. 
Conversely, if such pathways are absent or involve very high-energy states, the thermalization temperature may be substantially higher.

We simulate neutron capture nucleosynthesis with the Portable Routines for Integrated nucleoSynthesis Modeling (PRISM) astrophysical reaction network \cite{Sprouse21}, which has recently been updated to include support for nuclear excited states. 
Internal transition rates between the ground and isomeric states are computed using the work of \cite{Misch21}; nuclear $\beta$-decay transitions between states are also calculated with this machinery.
We calculate the population of excited states from neutron capture using the TALYS statistical Hauser-Feshbach code \cite{Koning2023}. 
Both of these tools use information on nuclear levels provided by the Evaluated Nuclear Structure Data Format (ENSDF). 
We use the trajectories from recent work to simulate the conditions of both nucleosynthesis processes. 
A weak $r$ process that makes it out to the second peak and an $i$ process is considered; further details about the trajectories can be found in Ref.~\cite{Mumpower25}. 

\begin{figure}[t!]
\centering
 \includegraphics[width=\columnwidth]{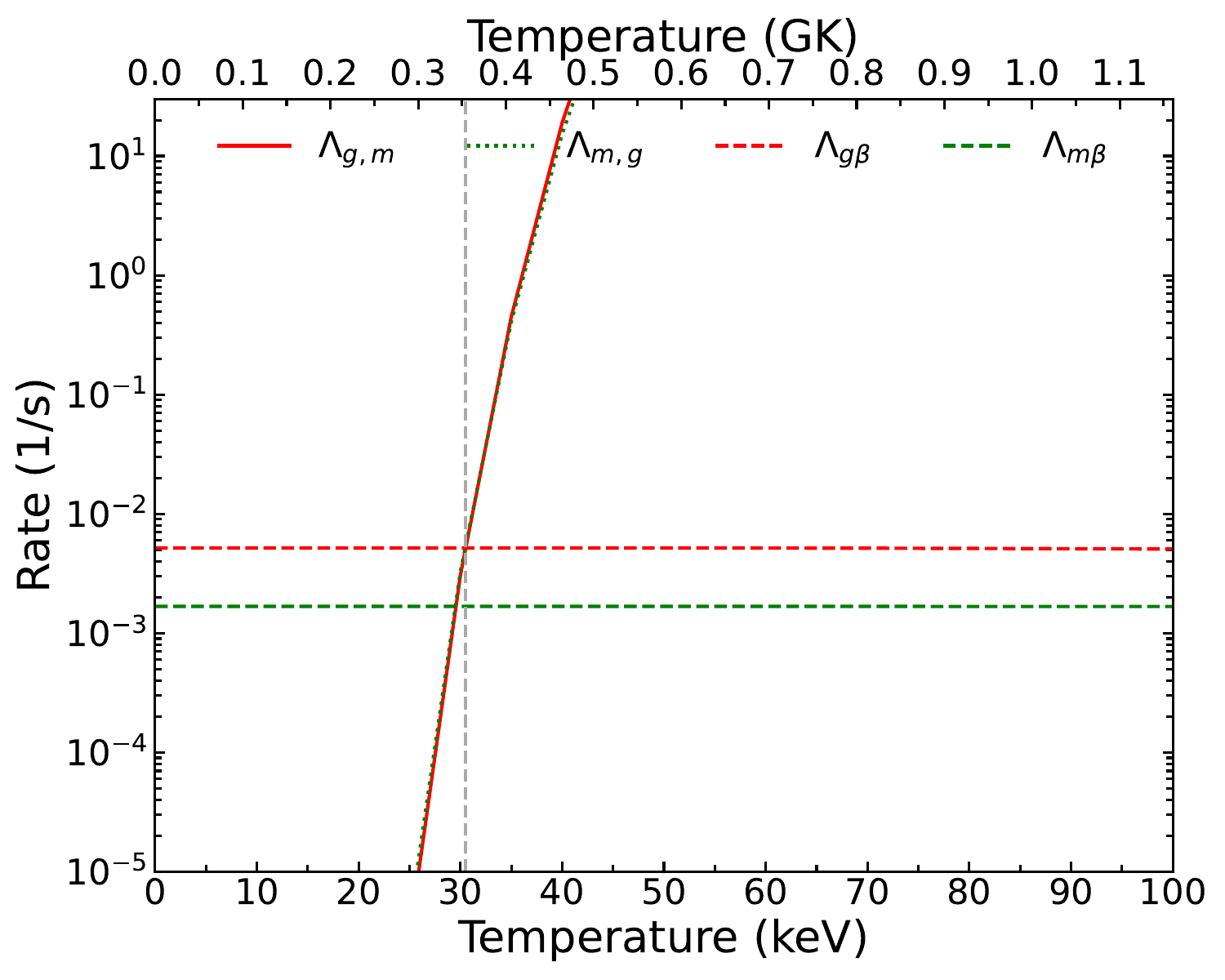}
 \caption{Astromer diagram for $^{129}$Sn. The solid line indicates transition from the ground state to isomeric state and the dotted line indicates the reverse. Rates of $\beta$-decay for the two states are shown with dashed lines. With present data the $\beta$ decay thermalization temperature, indicated by the vertical grey dashed line, is estimated to be 30 keV, below which the nucleus may need to be treated as two separate species in astrophysical reaction networks. 
 \label{fig:129Sn}}
\end{figure}

The relevant transitions associated with the $^{129}$Sn nucleus are shown in Fig. \ref{fig:129Sn}. 
This `astromer diagram' indicates the effective transitions from ground to isomer (solid red), from isomer to ground (dotted green), and the rates of $\beta$ decay for the two states (dashed lines). 
The last overlap of the transition rate with the $\beta$-decay rate of the respective rate defines the thermalization temperature, shown with a vertical grey dashed line. 
Above the thermalization temperature (here with respect to $\beta$-decay), the nucleus may be treated as one species. 
Once the temperature falls below this value, the two states must be treated as separate species in astrophysical reaction networks as the transition rate between nuclear states competes with the $\beta$-decay rate of a given state. 
Below this critical threshold, the excited state may now act as an astromer \cite{Misch21}. 
In the case of $^{129}$Sn, the thermalization temperature is predicted to be around 30 keV, set by the measurements discussed above. 

\begin{figure}[t!]
\centering
 \includegraphics[width=\columnwidth]{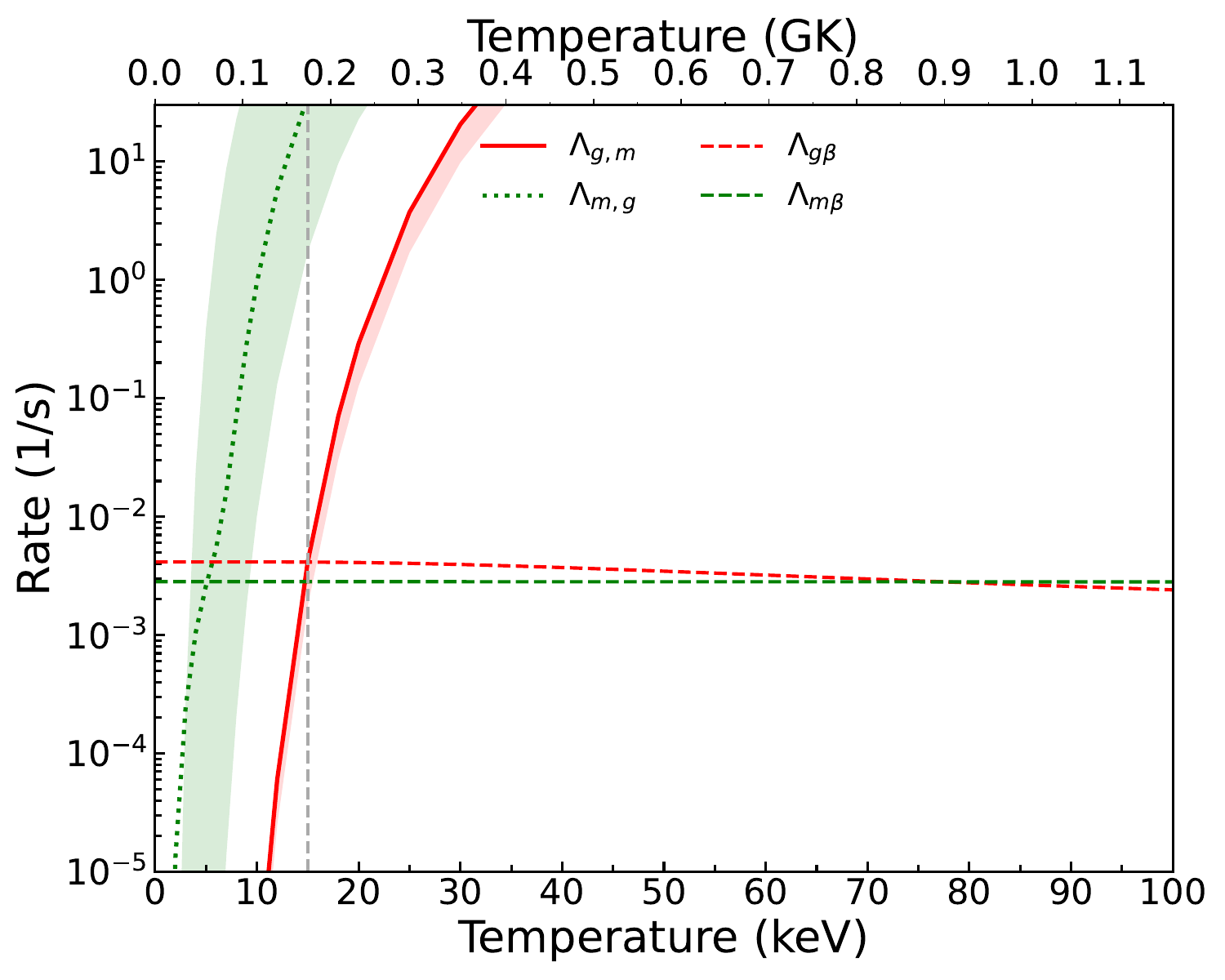}
 \caption{Astromer diagram for $^{132}$Sb. Lines are the same as in Fig. \ref{fig:129Sn}, with green shaded band indicating uncertainty from 50 keV uncertainty in the AME. With present data the $\beta$-decay thermalization temperature is estimated to be 15 keV, below which the nucleus may need to be treated as two separate species in astrophysical reaction networks. 
 \label{fig:132Sb}}
\end{figure}

\begin{figure}[t!]
\centering
 \includegraphics[width=\columnwidth]{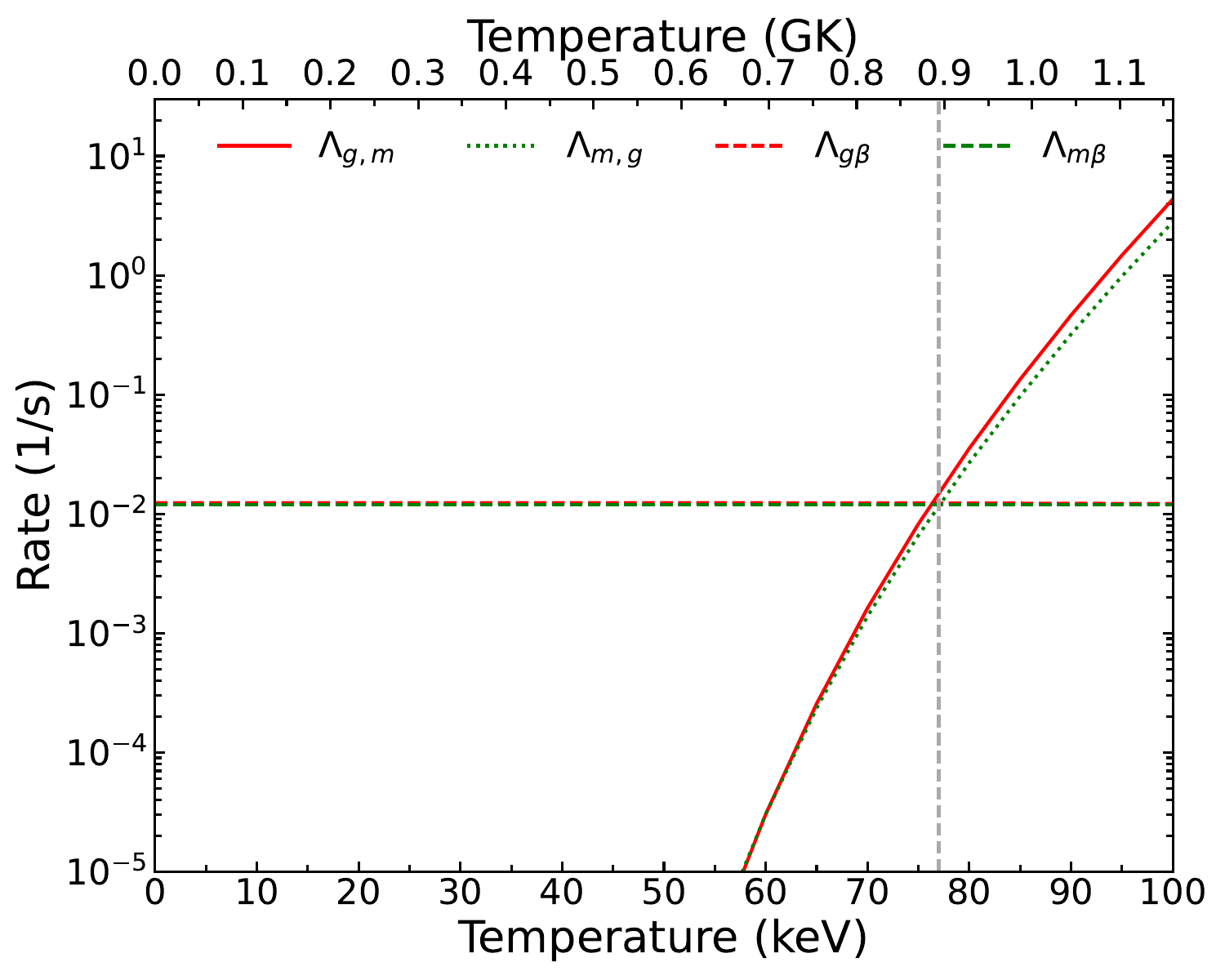}
 \caption{Astromer diagram for $^{131}$Sn. Lines are the same as in previous two figures. With present data the $\beta$-decay thermalization temperature is estimated to be 77 keV, below which the nucleus may need to be treated as two separate species in astrophysical reaction networks. 
 \label{fig:131Sn}}
\end{figure}

Our measurements yield an estimate of 15 keV for the thermalization temperature of $^{132}$Sb, as seen in the astromer diagram in Fig. \ref{fig:132Sb}. 
The new mass measurements take the previous uncertainties, as indicated by the shaded regions, down to the precise lines now shown. 
In the case of $^{131}$Sn, the thermalization temperature is  higher, around 77 keV (See astromer diagram, Fig. \ref{fig:131Sn}). 
The high thermalization temperature estimate could come down with further measurement of additional nuclear levels. 
Nonetheless, since the $\beta$-decay rates of the two states are nearly identical, the impact of populating the excited state is moderated. 

\begin{figure}[t!]
\centering
 \includegraphics[width=\columnwidth]{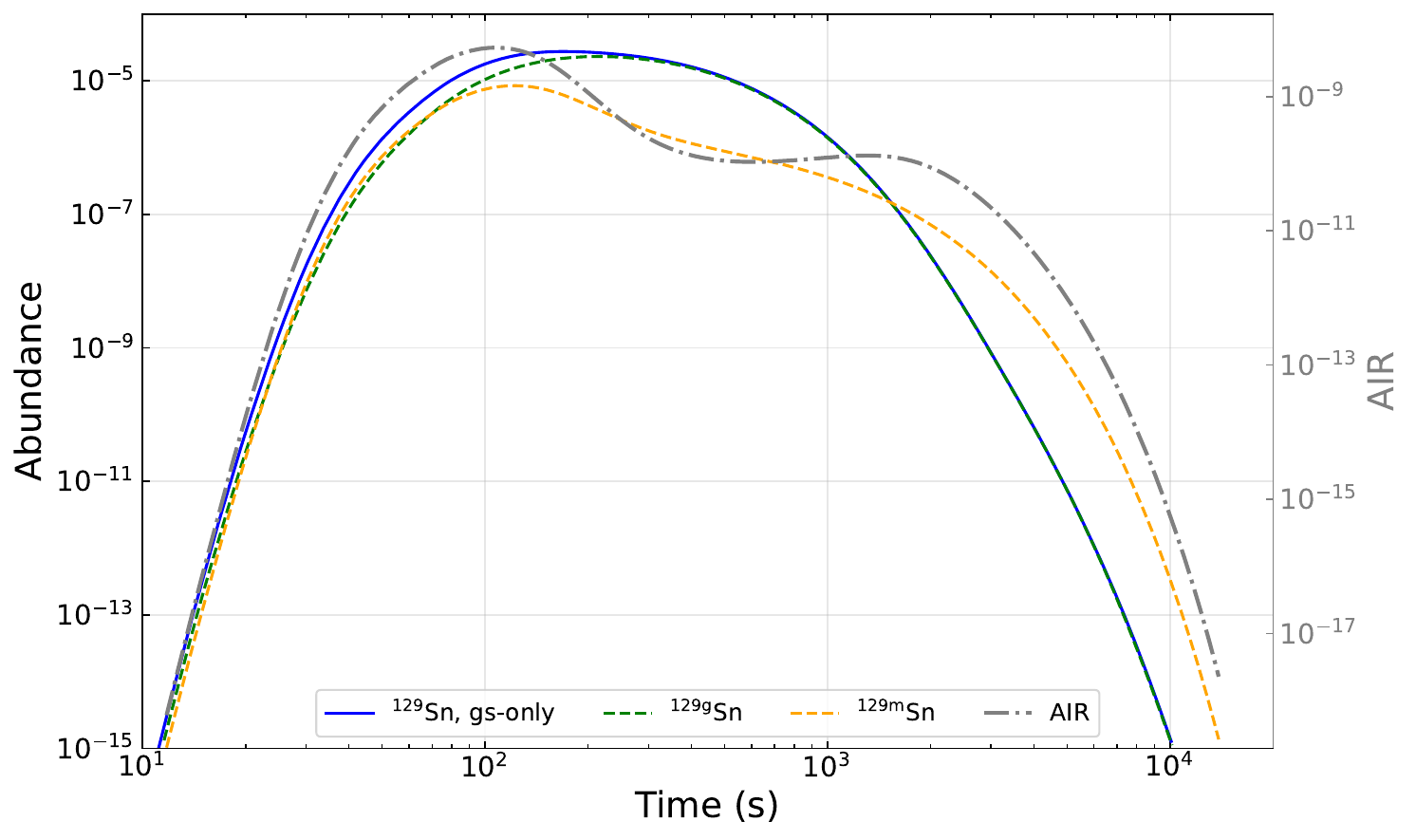}
 \caption{A network calculation showing the behavior of $^{129}$Sn in the astrophysical $i$-process. When treated as a single nucleus (solid line) $^{129}$Sn has a larger population. When the ground state is treated separately from the isomeric state, $^{129m}$Sn is found to be an astromer at early and late times (when the green curve is above orange). This can be verified by the AIR metric (dash dotted line read from the right Y axis).
 \label{fig:iprocess129Sn}}
\end{figure}

\begin{figure}[t!]
\centering
 \includegraphics[width=\columnwidth]{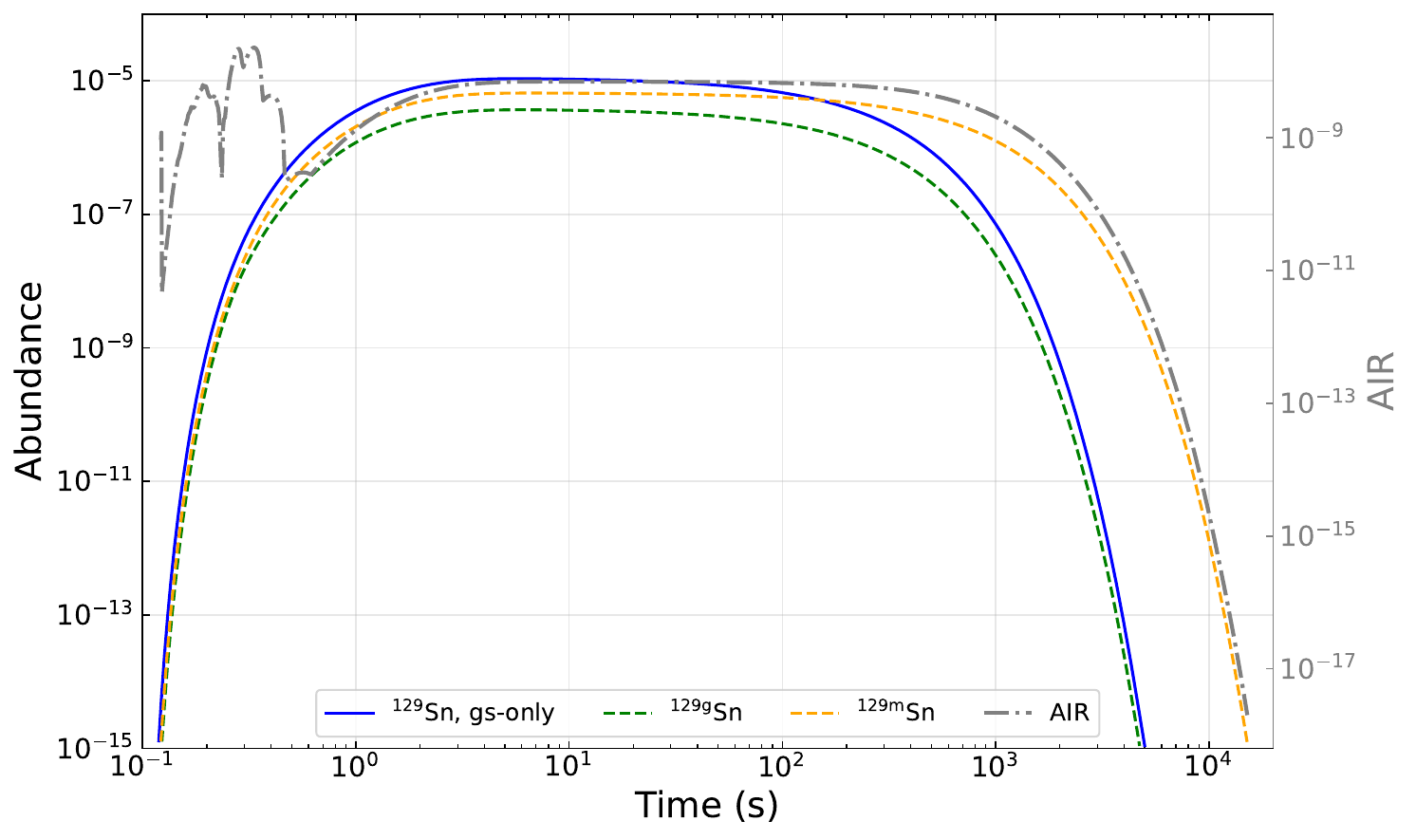}
 \caption{A network calculation showing the behavior of $^{129}$Sn in the astrophysical $r$-process. The solid line indicates $^{129}$Sn treated as a single nucleus. When the ground state is treated separately from the isomeric state, $^{129m}$Sn is found to be an astromer at all times, as indicated by the AIR metric (dash dotted line read from right Y axis)
 \label{fig:rprocess129Sn}}
\end{figure}

We now discuss the behavior of $^{129}$Sn, $^{131}$Sn, and $^{132}$Sb in network calculations of $i$-process and $r$-process nucleosynthesis. 
In the $i$-process, $^{129}$Sn is found to be an astromer at both early and late times during the duration of neutron capture. 
Treating this nucleus and others in the network as distinct isomers and ground states reveals the more complex behavior that is otherwise hidden when they are combined into a single species. 
This behavior is shown in Fig. \ref{fig:iprocess129Sn}. 
In several astrophysical conditions conducive to $r$-process nucleosynthesis \cite{Mumpower25}, $^{129}$Sn is found to be an astromer at all times; the isomeric state always more populated than the ground state. 
Fig. \ref{fig:rprocess129Sn} shows the behavior of $^{129}$Sn in the $r$-process. 
In both the $i$-process and $r$-process simulations, the abundance evolution for $^{129}$Sn is altered compared to the ground state only case. 
Such deviations have the potential to alter observable signals, e.g. in the context of kilonova \cite{Fujimoto20}, by changing the timing of energy release in the decay back to stability and thus altering the light curves. We defer a detailed analysis of these cumulative effects to future work.

In the case of $^{132}$Sb, the population of the isomeric state is negligible under both astrophysical scenarios considered here. 
Radiative capture on the ground state of $^{131}$Sb favors the direct population of the $^{132}$Sb ground state, while production of the isomeric state is consistently suppressed at all temperatures. 
Moreover, the presence of nearby short-lived states at excitation energies of $E_x = 85.55$ keV and $E_x = 254.5$ keV  \cite{ENSDF} facilitates rapid depopulation of the isomer, even at modest temperatures. 
Consequently, the isomeric state does not play a significant role in the nucleosynthesis of $^{132}$Sb. 
For the conditions studied in this work the reaction flow proceeds predominantly through the ground state. 

\begin{figure}[t!]
\centering
 \includegraphics[width=\columnwidth]{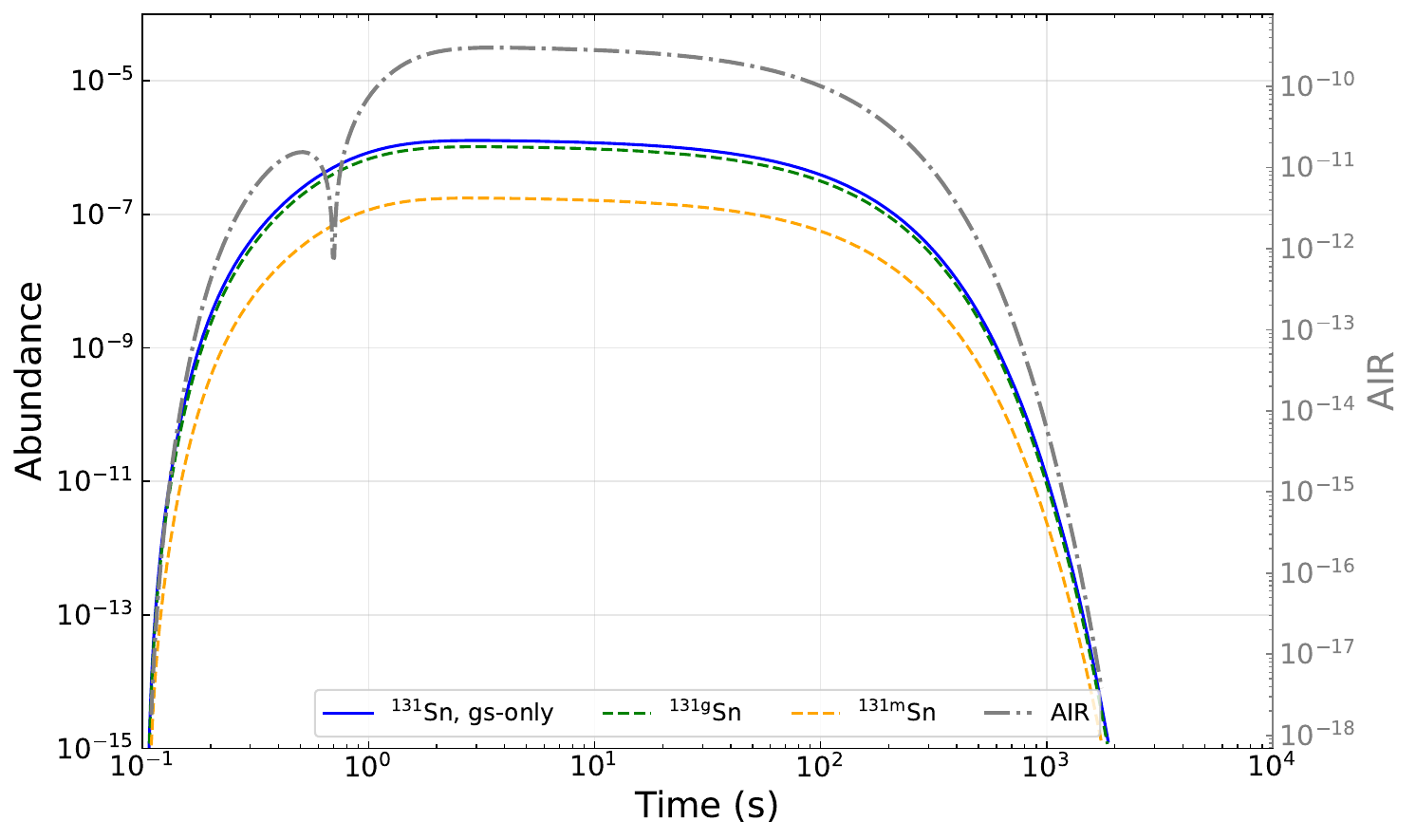}
 \caption{A network calculation showing the behavior of $^{131}$Sn in the astrophysical $r$-process. The solid line indicates $^{131}$Sn treated as a single nucleus. When the ground state is treated separately from the isomeric state, $^{131m}$Sn is found to be an astromer, as indicated by the AIR metric (dash dotted line read from right Y-axis). 
 \label{fig:rprocess131Sn}}
\end{figure}

The case of $^{131}$Sn exhibits behavior intermediate to that of $^{129}$Sn and $^{132}$Sb. 
Figure \ref{fig:rprocess131Sn} shows the population of the states in $^{131}$Sn. 
The maximum relative population ratio of the isomer to the ground state peaks at approximately 15\% in this $r$-process simulation, compared to roughly 12\% in the $i$-process. 
Feeding into the isomeric state occurs primarily through thermal transitions and $\beta$-decay. 
The nucleus $^{131}$Sn can be populated through two potential radiative neutron capture pathways: one proceeding from the ground state of $^{130}$Sn, and the other from its excited state at 1946 keV \cite{NDS130}. 
However, the relatively high excitation energy of the latter state suppresses its contribution, meaning the production of $^{131}$Sn occurs predominantly through capture on the $^{130}$Sn ground state. 
Furthermore, due to the large spin difference involved, neutron capture preferentially populates the ground state of $^{131}$Sn. 
Nevertheless, the continual population of the excited state in $^{131}$Sn—despite remaining consistently lower than that of the ground state—exhibits astromeric behavior. 
This is evident in the trend of the AIR. At early times, the nucleus is relatively thermalized and the AIR is small but non-zero. 
As thermal equilibrium breaks down, the balance between the thermal populations of the isomer and the ground state shifts, causing a cusp in the AIR at approximately 0.8 seconds. 
Once $\beta$-decay becomes the primary population channel, depopulation of this state via transitions to the ground state cannot keep pace, and the state becomes an astromer. 
Despite the existence of this astromer, the final evolution of the total abundance flowing through $^{131}$Sn remains identical to the ground-state-only scenario. 
Consequently, the impact on observable signals is likely minimal.

\section{Conclusions}
In conclusion, we have conducted new, high-precision mass measurements of $^{129g,m}$Sn, $^{131g,m}$Sn, and $^{132g,m}$Sb using the Canadian Penning Trap at Argonne National Laboratory's ATLAS user facility's CARIBU facility. Our results for these ground and isomeric states are in good agreement with most of the preceding literature values, and provide the highest-precision determination of the excitation energies of these long-lived isomeric states directly from their masses. We then used these new excitation energies to determine the effective thermal transitions between the ground and isomeric states, their relative $\beta$ decay rates, and the  associated cross sections into and out of these states. The excitation energies were also used to estimate the thermalization temperature with respect to $\beta$-decay, above which these nuclei may be correctly treated in reaction networks as one species. 
Reaction network calculations were used to establish that $^{129m}$Sn must be treated as an astromer in the $i$-process and $r$-process. 
The excited state in $^{131}$Sn may need to be treated as an astromer depending on the astrophysical conditions. 
The state in $^{132}$Sb was determined not to behave as astromers under the conditions studied in this work. 
As the thermalization of these nuclei is dependent on the excited states above the isomers, a full understanding of their astromeric nature would benefit from detailed spectroscopic studies of the spins and parities of the known states, a search for potential missing states, and on the determination of their feeding via $\beta$ decays. 

\begin{acknowledgments} 
This work was performed with the support of U.S. Department of Energy, Office of Science, Office of Nuclear Physics under Contracts No. DE-AC02-06CH11357 (ANL), DE-AC52-07NA27344 (LLNL), and DE-AC02-05CH11231 (LBNL);  the Natural Sciences and Engineering Research Council of Canada under Grant No. SAPPJ-2018-00028; and the US National Science Foundation under Grant No. PHY-2310059. LANL is operated by Triad National Security, LLC, for the National Nuclear Security Administration of U.S. Department of Energy (Contract No. 89233218CNA000001). This research used resources of ANL’s ATLAS facility, which is a DOE Office of Science User Facility. 

AAV, WSP, RO and DEMH would like to acknowledge the contributions of SKB and MBC. Research reported in this publication was supported by the U.S. Department of Energy LDRD program at Los Alamos National Laboratory (Project Nos. ER20230052ER and 20240004DR). At LLNL, this work was supported in part by the U.S. DOE NNSA through the Office of Nonproliferation Research and Development (NA-22) under the Funding Opportunity Announcement LAB 17-1763.
\end{acknowledgments}

\bibliographystyle{apsrev4-2}
\bibliography{AstromersSnSb} 

\end{document}